\documentclass{ws-mpla}

\usepackage{latexsym}

\begin{document}

\markboth{E. M. Drobyshevski}
{Low-energy recoil ion channeling in NaI(Tl) and DM detection}
\title{CHANNELING EFFECT AND IMPROVEMENT OF THE EFFICIENCY OF CHARGED 
PARTICLE REGISTRATION WITH CRYSTAL SCINTILLATORS}

\author{E.M.Drobyshevski}
\address{Ioffe Physical-Technical Institute, Russian Academy of Sciences, 194021 St.Petersburg, Russia\\
emdrob@mail.ioffe.ru}

\maketitle

\begin{abstract}
Energy of ions (recoil nuclei) channeled along definite directions in 
crystals is transferred to the lattice electrons mainly. In NaI(Tl)-type 
scintillators, this leads to increasing the light yield from $\sim $10{\%} 
to $\sim $100{\%} when compared with the yield for electrons of the same 
energy. Taking into account this effect at processing data of DAMA/NaI 
experiments in Gran Sasso, which had demonstrated the year modulation of 
number of signals in a range of 2-6 keV of electron equivalent, reveals that 
DAMA/NaI results could be caused by $\sim $6$\times $10$^{-7}$ cm$^{-2}$ 
s$^{-1}$ flux of daemons (Dark Electric Matter Objects -- presumably 
Planckian relic particles) falling out from strongly elongated heliocentric 
orbits with velocities of 30-50 km/s. The flux value and the 2-6 keV signal 
intensity agree rather well with values emerging from our former estimates 
and interpretations of ground-level and underground measurements.

\end{abstract}

\ccode{PACS Nos.: 29.40.Mc, 61.85.+p, 95.35.+d}

\section{Introduction}
Experiments aimed at the detection of dark matter (DM) objects and the 
determination of their properties by investigating the recoil ions (nuclei) 
with scintillation detectors rest essentially on the knowledge of the 
efficiency parameter $q$ (quenching factor), the light yield ratio for the 
recoil ion and the electron of the same energy.$^{1}$ Our interest in this 
parameter stems from the possibility that the year modulation in the number 
of signals revealed with a high confidence in the DAMA/NaI(Tl) and 
DAMA/LIBRA search for WIMPs (Weakly Interacting Massive Particles, 
hypothetical neutral particles $\sim $10$^{2}$ GeV in mass considered to be 
candidates for DM)$^{2,3}$ can be assigned to registration of a flux of 
electrically negative daemons (Dark Electric Matter Objects, presumably the 
elementary Planckian black holes, $\sim $10$^{19}$ GeV in mass and with an 
electric charge \textit{Ze} $\approx $ 10$e)$ captured out of the galactic disk by 
combined action of the moving Sun and of the Earth into strongly elongated, 
Earth-crossing heliocentric orbits (SEECHOs). These orbits crowd in the 
antapex zone beyond the Sun, which is crossed by the Earth sometime in early 
June. It is essential that for $q \approx $ 1 the amplitude range of the 
significant 2-6-keV signals in the DAMA/NaI(Tl) and DAMA/LIBRA experiments 
coincides with the energy of the iodine recoil ions if the latter are 
knocked out elastically by daemons falling on the Earth from SEECHOs with 
velocities of 30-50 km/s.$^{4,5}$

The parameter $q$ for the I (and Na) ions is determined traditionally in 
neutron scattering experiments. These are either direct measurements of the 
amplitude of the scintillations correlating with monoenergetic neutrons 
scattered to a preset angle$^{6-9}$ or statistical treatment (by Monte-Carlo 
simulation) of the scintillation spectrum of a NaI(Tl) crystal irradiated by 
neutrons with an average energy of 2.35 MeV from a $^{252}$Cf 
source.$^{10,11}$ In the first case, $q_{Na}$ = 25-30{\%} for Na ions with a 
recoil energy $E_{r} >$ 4 keV and $q_{I}$ = 8-8.6{\%} for $E_{r} >$ 10 keV. 
In the second case, $q_{Na}$ = 30-40{\%} for $E_{r} >$ 5 keV and $q_{I}$ = 
5-9{\%} for $E_{r} >$ 22 keV (see Table 1). We see that the data obtained 
by different groups fit fairly well.

Such calibrations of NaI(Tl) scintillators are usually assumed to mean that 
$q$ does not depend on the energy of the recoil ion $E_{r}$ ($i.e$., $\partial 
q$/$\partial E_{r}$ = 0, see Refs. 7,12). Actually, while this conclusion 
does not follow from anything else, it is buttressed by the observation that 
the light energy yield of an inorganic scintillator (for NaI(Tl) it amounts 
to about 15{\%}$^{1})$ depends only weakly on electron energy, varying by 
less than 20{\%} from 5 to 1000 keV,$^{1,13,14}$ just as the light energy 
yield for heavy particles with $E >$ 1 MeV. On the other hand, as Birks$^{1}$ 
pointed out long ago, $q >$ 1 in heavy metal iodides for protons and 
deuterons with $E \approx $ 5 MeV. Davis \textit{et al}.$^{15}$ (see also Ref. 8) observed 
$q$ to grow for Ca and F ions with decreasing $E_{r}$ in the CaF$_{2}$(Eu) 
crystal. On the contrary, in liquid Xe the quenching relative to gamma rays 
was found by Aprile \textit{et al.}$^{16}$ to grow almost two-fold with increasing 
$E_{r}$ from 10.4 to 56.5 keV. So one cannot rule out the possibility 
that the conclusion of the constancy of $q$ in the particular case of NaI(Tl) 
is simply a consequence of the difficulties encountered in measurements in 
the keV range (even for electrons, measurements on NaI(Tl) were performed 
only down to 0.87 keV$^{17})$.

As we are going to show, there are grounds to believe that the value of $q$ for 
a keV-range iodine recoil ions in the NaI(Tl) crystalline scintillator may 
approach 1, and that under certain conditions the efficiency of 
scintillation excitation by ions may even exceed somewhat that by electrons.
From this follows the conclusion that DAMA/NaI results are quantitatively explicable within the 
framework of the daemon paradigm.

\section{On Interaction of keV-range Ions with a Solid Body}

\subsection{Amorphous solids}

When an ion enters a solid, it passes without stopping a certain path. This 
path length is determined by interaction with nuclei (ions) and the 
electronic component, with the major contribution at low energies coming 
from slowing down by nuclei, and at high energies, by electrons. The 
stopping power due to electrons is proportional to the ion velocity.

The first self-consistent theory of ion slowing down in monoatomic amorphous 
materials (including gases, but disregarding the ionization and dissociation 
of molecules) was developed by Lindhard, Scharff, and Schiott (LSS).$^{18}$ 
They divided the stopping power into two components associated, accordingly, 
with nuclei and electrons. Then the scintillator efficiency $q$ is determined 
simply as the ratio of the energy imparted to the electronic component to 
the total kinetic energy of the ion (i.e., to the sum of the nuclear and 
electronic stopping powers). It is this theory that is frequently compared 
with measurements associated with the motion of recoil ions in crystalline 
scintillators.$^{6,10,19}$ The theoretical LSS values of $q$ were shown to 
exceed its experimental values by approximately a factor 2-2.5.

Firsov$^{20}$ took into account the ionization resulting from overlap of the 
electronic shells of the moving and the target atoms, an approach that 
provides in some cases a better agreement with experiment than LSS 
theory.$^{21,22}$ Further progress in the field of theory involved primarily 
a closer analysis of the effective charge of the projectile ion through a 
consecutive refinement of the quantum-mechanical description of ion 
interaction with the target atoms, a point of particular importance below 
the Bohr velocity. Ziegler$^{23}$ outlined the history and present status of 
the problem involving the motion of ions with $E \approx $ 1 keV-1000 MeV in 
amorphous solids, and developed a computer code SRIM, which permits 
calculation of the ion range and scattering, as well as the ion and electron 
stopping powers for any combination of substances, including nonmonoatomic 
targets. Compounds are analyzed using the Bragg rule that states that the 
stopping power of an ion in a compound can be approximated by a combination 
of stopping powers of the constituent target ions. This rule was shown to 
hold the better, the higher are the atomic numbers of the elements. SRIM 
calculations suggest that the mean path length to rest in amorphous NaI of a 
2-keV iodine ion projected on the initial direction of motion is \textit{$\lambda $}$_{a}$ = 50 
{\AA}, a comparison of the ion with electron stopping powers yields $q$ = 
0.068, and further on (see Table 1). We readily see that the values of $q$ 
following from the nowaday theory$^{23}$ agree fairly well with experiment 
(see Sec. 1 above and Refs. 6-11).

\begin{table}[t]
\begin{center}
\tbl{Comparison of the experimentally determined quenching factors $q$ for 
Na and I recoil ions in NaI(Tl) scintillator ($q_{exp}$ and $ q_{exp+MC}$ -- 
the both determined with making use the neutron scattering technique, but the $ q_{exp+MC}$ 
is a result of statistical treatment; see text) with the SRIM-code 
calculated values of $ q$ (and \textit{$\lambda $}$_{a}$ -- the mean path length to rest) in 
amorphous NaI.}
{\begin{tabular}{|c|c|c|c|c|c|c|c|}
\hline
\multicolumn{3}{|c|}{$E_{r}$, keV} & 
2& 
4& 
10& 
20& 
50 \\
\hline
\raisebox{-4.50ex}[0cm][0cm]{Na}& 
\multicolumn{2}{|c|}{$q_{exp}$} & 
& 
\multicolumn{4}{|l|}{0.25-0.30 (for $E_{r} >$ 4 keV)}  \\
\cline{2-8} 
 & 
\multicolumn{2}{|c|}{$q_{exp+MC}$} & 
& 
\multicolumn{4}{||l|}{0.3-0.4 (for $E_{r} >$ 5 keV)}  \\
\cline{2-8} 
 & 
\raisebox{-1.50ex}[0cm][0cm]{S \par R \par I \par M}& 
$q$& 
0.153& 
0.181& 
0.236& 
0.301& 
0.434 \\
\cline{3-8} 
 & 
 & 
$\lambda _{a}$, \AA& 
71& 
115& 
233& 
421& 
985 \\
\hline
\raisebox{-4.50ex}[0cm][0cm]{I}& 
\multicolumn{2}{|c|}{$q_{exp}$} & 
& 
& 
\multicolumn{3}{|l|}{0.080-0.086 (for $E_{r} >$ 10 keV)}  \\
\cline{2-8} 
 & 
\multicolumn{2}{|c|}{$q_{exp+MC}$} & 
& 
& 
& 
\multicolumn{2}{||l|}{0.05-0.09 (for $E_{r} >$ 22 keV)}  \\
\cline{2-8} 
 & 
\raisebox{-1.50ex}[0cm][0cm]{S \par R \par I \par M}& 
$q$& 
0.068& 
0.070& 
0.072& 
0.085& 
0.103 \\
\cline{3-8} 
 & 
 & 
$\lambda _{a}$, \AA& 
50& 
70& 
114& 
170& 
307 \\
\hline
\end{tabular}}
\end{center}
\end{table}

\subsection{The channeling effect in crystals}

Motion of an ion in crystals, in particular, in inorganic scintillators, 
differs substantially in many respects from that in amorphous solids. The 
ordered arrangement of nuclei in crystals makes possible the so-called 
channeling of ions propagating along certain directions (crystallographic 
axes and planes). Channeling becomes manifest in an anomalously deep 
penetration of ions into a target, an effect discovered half a century ago 
by Bredov and Okuneva.$^{24}$ They observed penetration of 4-keV 
$^{134}$Cs$^{+}$ ions into a Ge crystal to a depth \textit{$\lambda $}$_{c} \sim $ 10$^{3}$ {\AA} 
(to feel the difference, a 4-keV Cs$^{+}$ ion would penetrate into amorphous 
Ge, according to SRIM calculations, only to a depth \textit{$\lambda $}$_{a}$ = 44 {\AA} for 
the ion to electron stopping power ratio in an amorphous solid \textit{$\kappa $} = 32). The 
explanation for such a deep penetration, as was found 6 years after the 
discovery of Bredov and Okuneva, lies in the ion stopping power decreasing 
strongly along some directions in a crystal (see review Ref. 22). Within a 
channel, stopping is dominated by electrons. This is seen already from the 
above comparison of the data of Bredov and Okuneva with calculations made 
for amorphous Ge (whence it follows, in particular, that the mean path 
length of an ion in the channel, \textit{$\lambda $}$_{c}$, exceeds that in an amorphous solid 
by a factor $\sim $(\textit{$\kappa $} + 1), i.e., \textit{$\lambda $}$_{c} \approx $ (\textit{$\kappa $} + 1)\textit{$\lambda $}$_{a}$ = \textit{$\lambda $}$_{a}$/$q)$.

Measurements of the mean range of different ions, say, in W crystal likewise 
produce remarkable results. For instance, in the case of Xe$^{+}$ 
non-channeled ions, the ion stopping power, which originally is higher than 
that of the electronic one, decreases to become equal to the electronic one 
(which remains proportional all the time to the ion velocity) only at 
2.7-MeV ion energy, whereas the ion stopping power observed under channeling 
conditions becomes equal to the electronic one already at 4 keV.$^{21}$

The energy of a channeled ion is transferred primarily to electrons, both 
single and to their continuum. The electrons transform eventually a part of this 
energy into light, as if the scintillator was irradiated by electrons from 
the outset. So it is obvious that even this one point leads to $q \to $ 1 for 
the channeled ion case.

This could be a convenient time to stop discussion of the channeling effect 
and of its contribution to high-efficiency detection of low-energy recoil 
ions, were it not for some implications which we shall consider below.

\subsection{An influence of making and breaking of 
quasi-molecular bonds of channeled ions with ions 
in the channel walls}

Another process capable of improving the efficiency of scintillation 
radiation generated by the channeled recoil ions is the consecutive 
formation by them of quasi-molecular bonds with channel wall atoms resulting 
from the overlap of their electronic shells (while the corresponding theory 
is still lacking, recalling the concepts of Firsov$^{20}$ would be here 
appropriate). The energy of dissociation of a single NaI ionic molecule is 
3.16 eV, and the electron affinity of the iodine atom is 3.06 eV, which 
suggests that the particles knocked out from the NaI lattice are negative 
I$^{-}$ recoil ions (recall that the iodine ionization energy is 10.45 eV, 
and the average energy required to remove an ion from the crystal lattice is 
20-30 eV). Therefore, it appears reasonable to assume that consecutive 
ruptures of these bonds initiated by continuous motion of the ion would 
ideally release in the form of photons an energy of $\sim $2.5-3 eV per each 
2 {\AA} (the distance between like ions along the channel in the NaI 
lattice). If there were no other channels and mechanisms of energy loss, a 
4-keV ion would expend its energy in a distance of $\sim $3000 {\AA}. The 
existence of other ways of energy loss in interaction with the electronic 
component reduces naturally this path to one third, so that this mechanism 
would boost the efficiency of transformation of the ion's kinetic energy to 
light by 30{\%}, to result in $q >$ 1. We thus see that the above process, 
even if it were realized only partially, would give rise to a marked 
increase in the efficiency of ion energy transformation to the scintillation 
radiation of the crystal.

\subsection{Dechanneling due to collisions with 
Tl$^{+}$ luminescent centers in NaI(Tl)}

Massive NaI crystals used as scintillators are not perfect, if only because 
the conditions of their growth are not ideal. Their lattice contains 
inevitably various defects in the form of dislocations and so on. As it is 
well known,$^{21,22,25}$ presence of defects and thermal vibrations in 
lattice are among the main causes responsible for the ion dechanneling. The 
main source of defects in the NaI(Tl) lattice are thallium atoms 
intentionally doped into the crystal to create luminescent Tl$^{+}$ centers, 
which increase substantially the light yield of the phosphor.$^{1}$ The 
optimum molar concentration of Tl in NaI(Tl) is 0.0013.$^{1}$ Assuming the 
radius of Tl$^{+ }$ions to be $\sim $1.5 {\AA}, and that of I$^{-}$ ions, 
$\sim $2.2 {\AA}, we obtain $\sim $1200 {\AA} for the mean path length of an 
I$^{-}$ ion between consecutive collisions with the ions of the Tl$^{+}$ 
centers. This length corresponds to \textit{$\lambda $}$_{a}$ = 86 {\AA} for the iodine ion 
with $E_{r}$ = 6 keV, with $q$ = 0.072, which yields exactly \textit{$\lambda $}$_{c}$ = \textit{$\lambda $}$_{a}$/$q$ = 
1200 {\AA}. Thus, starting already from an energy $\sim $6-10 keV the iodine 
recoil ions will undergo dechanneling in NaI(Tl) because of their 
interaction with the heavy ($A$ = 204.4, $Z_{n}$ = 81) Tl$^{+}$ ions mainly 
(note that ion dechanneling is frequently used to probe the position of 
foreign atoms in crystals$^{25})$. It is conceivable that dechanneling of 
this kind in NaI(Tl) in collisions with Tl$^{+}$ luminescent centers is 
accompanied by an excess scintillation light yield (i.e., again by an 
increase of $q$ above 1). The dechanneling of ions due to their scattering on 
other type defects has a lower probability owing to their significantly 
lower concentration.

\section{Probability of the Recoil Ion Channeling and the Near-Earth Daemon 
Flux as Observed by DAMA/NaI}

The recoil ion (nucleus) knocked out of a lattice site by a heavy (DM) 
particle moves, generally speaking, in an arbitrary direction. The 
probability of channeling for an ion crossing a crystallographic channel 
(axial or planar) depends on the angle between its trajectory and the 
channel direction and is inversely proportional to the quadruple root of its 
energy for an axial channel and to the square root (i.e. to its velocity) 
for a planar one. We used recommendations of Appleton \textit{et al}.$^{25}$ to calculate 
the angles of entry into the principal accessible channels, axial and 
planar, for the NaI crystal (NaCl-type cubic crystal, lattice constant 6.473 
{\AA}). For singly charged, 4-keV iodine ions the critical angle for the 
$<$100$>$ axial channels is 6.4$^{o}$, for the $<$110$>$ channels it is 
4.9$^{o}$, and for the $<$111$>$ channels, 2.8$^{o}$; for the planar 
{\{}100{\}} channels it is 4.1$^{o}$, for {\{}110{\}} it is 3.2$^{o}$, and 
for {\{}111{\}}, 2.7$^{o}$. Summing up the solid angles formed by these 
channels we obtain that the channeling probability for a 4-keV iodine recoil 
ion ejected within a crystal in an arbitrary direction is in this particular 
case as high as nearly 20{\%}. Because, as we have seen, for channeling 
recoil ions $q \approx $ 1, it thus follows that the efficiency of detection 
of these ions, while being naturally somewhat lower, will nevertheless 
amount to about 20{\%}. 

Assuming now that the DAMA/NaI and DAMA/LIBRA experiments$^{2,3}$ detect 
with an efficiency \textit{$\eta $} = 20{\%} (for the remaining 80{\%} unchanneled ions $q$ 
$\approx $ 0.1, because for the 2-keV detection threshold in this experiment 
these ions pass undetected) the flux of daemons crossing the Earth with $V$ = 
30-50 km/s in the antapex relative to the Sun region,$^{5}$ it is of 
interest to estimate this flux from available measurements (some possible 
processes of daemon interaction with matter were discussed somewhere 
earlier, see Refs. 4,5,26-28 for examples).

The double modulation amplitude (swing) of the number of events with a 
one-year period measured in these experiments is about 0.04 cpd/kg/keV in 
the 2-6-keV interval.$^{2,3}$ In the DAMA/NaI case the NaI(Tl) crystal mass 
was 96 kg, the crystal packing density was about 50{\%}, which yields for 
the effective cross section of the system about 1500 cm$^{2}$. Whence for 
the average 4-keV signal we obtain that the flux of SEECHO daemons falling 
on the Earth varies during a year from zero to $f_{\oplus } \approx $ 
(0.04$\times $96$\times $4)/(86400$\times $1500$\times $\textit{$\eta $}) = 6$\times 
$10$^{-7}$ cm$^{-2}$ s$^{-1}$ in June, a figure in a good agreement with our 
early estimates of the SEECHO flux of 3$\times $10$^{-7}$ cm$^{-2}$s$^{-1}$ 
(Ref. 26), which, in its turn, is not at odds with our surface-level$^{27}$ 
and underground$^{28}$ measurements of the flux of daemons with $V$ = 10-15 
km/s, which fall on the Earth near equinoxes from near-Earth, almost 
circular heliocentric orbits. Similar flux of daemons is needed to explain 
``Troitsk anomaly'', - viz. a periodic drift of the tritium $\beta 
$-spectrum tail position in experiments on direct neutrino mass measurement 
exploiting the extended T$_{2}$ gas source with Nb-containing 
superconducting magnetic coils (for more details see Ref. 29 and references 
therein).

\section{Conclusions}

One may thus conclude that channeling should play a dominant role in 
measurement of low-energy (keV-range) recoil ions with crystal detectors of 
the NaI(Tl) type. This effect reduces strongly nuclear stopping, to make 
interaction with valence electrons of the crystal lattice, including 
formation and rupture of quasi-molecular bonds with crystal atoms, the 
dominant factor in stopping of recoil ions. As a result of this highly 
efficient interaction, the light yield of the crystal may reach and even 
exceed that observed when the crystal is acted upon by electrons of the same 
(keV-range) energies; indeed, in the latter case electrons expend part of 
their energy also in excitation of the inner-shell atomic electrons.

At high energies (above, say, ten keV), besides the above-mentioned lower 
probability of entering into a channel due to the acceptance angle 
diminution, the channeled ion moving in the NaI(Tl) crystal leaves 
eventually the channel, if for no other reason than as a result of 
interaction with a Tl$^{+}$ center, well before exhausting its energy owing 
to the electron stopping power action (which, in particular, stresses the 
need of reducing and optimizing the activator content in the scintillator). 
The ion scattered out of the channel loses its energy primarily through 
nuclear stopping, and this is what accounts for quenching of the light 
yield. This is why, in particular, the efficiency of scintillation 
excitation in NaI(Tl) by iodine recoils with an energy of tens of keV is an 
order of magnitude lower than that by electrons, a point substantiated by 
neutron beam calibrations.$^{6-11}$

The above stresses the need for developing methods to directly calibrate 
crystal scintillators with keV-range ion beams, as this was done by Bredov 
and Okuneva$^{24}$ in the first semiconductor implantation experiments.

A recent publication by the EDELWEISS Collaboration$^{30}$ reports on 
neutron scattering measurements of the dependence of heat and ionization 
quenching factors in Ge on $E_{r}$ performed in the 10-100-keV range. As 
pointed out by the authors, as of today these are the most precise absolute 
measurements for any detectors employed in the direct DM search. In the 
above energy range, the ionization $q$ increases with increasing $E_{r}$ by a 
factor $\sim $1.7. On the other hand, as far back as 1971 Jones and 
Kraner$^{31}$ pointed out that for $E_{r} \le $ 1.8 keV, $q$ does increase 
too. An increase of $q$ for 2.7 $\le  E_{r} <$ 7-8 keV was reported by 
Messous \textit{et al}.$^{32}$ (see also Fig. 2 in Ref. 30). The latter authors ascribed 
the effect to the poor resolution of the detectors, whereas Jones and 
Kraner$^{31}$ explicitly suggest the possibility of strong channeling 
(\textit{sic}!) of recoil nuclei in the Ge crystal. Unfortunately, $E_{r }$= 10 keV for 
elastically knocked out Ge nuclei in the EDELWEISS experiment corresponds to 
the supermassive projectile velocity of $V \approx $ 80 km/s, a figure 
substantially in excess of the SEECHO daemon velocity of 50 km/s. Thus, we 
have to state once more regretfully that it is apparently only the DAMA/NaI and DAMA/LIBRA, 
besides our experiment, that are presently capable of detecting the fall of 
daemons from Earth-crossing orbits.

\section*{Acknowledgments}
The author expresses sincere gratitude to Prof. R. Bernabei (University 
``Tor Vergata'' and INFN, Rome) for attracting his attention to the problem 
of low conversion efficiency of the energy of recoil ions into that of 
scintillations, which has stimulated the present study. The author feels 
indebted also to Drs. I. T. Serenkov and V. I. Sakharov (Ioffe 
Physical-Technical Institute, RAS) for fruitful discussions of the 
channeling effects and corresponding calculations with the SRIM code.

\end{document}